\documentclass{elsart}

\usepackage{natbib} 

\usepackage{epsfig}

\usepackage{amssymb} 
\journal{New Astronomy}

\def\astrobj#1{#1}
\def\bibcode#1{}

\begin{document}

\runauthor{K.-H. Hofmann et al.}

\begin{frontmatter}

\title{Observations of Mira stars with the IOTA/FLUOR interferometer
       and comparison with Mira star models\thanksref{1}}
\thanks[1]{Based on observations collected at the IOTA/FLUOR interferometer, Whipple
Observatory, Mount Hopkins, Arizona.}
\author[a1]{K.-H. Hofmann}
\author[a1]{U. Beckmann}
\author[a1]{T. Bl\"ocker}
\author[a2]{V. Coud$\acute{\rm e}$ du Foresto}
\author[a3]{M. Lacasse}
\author[a2]{B. Mennesson}
\author[a3]{R. Millan-Gabet}
\author[a3]{S. Morel}
\author[a2]{G. Perrin}
\author[a3]{B. Pras}
\author[a2]{C. Ruilier}
\author[a1]{D. Schertl}
\author[a6]{M. Sch\"oller}
\author[a4]{M. Scholz}
\author[a5]{V. Shenavrin}
\author[a3]{W. Traub}
\author[a1]{G. Weigelt}
\author[a6]{M. Wittkowski}
\author[a5]{B. Yudin}
\address[a1]{Max-Planck-Institut f\"ur Radioastronomie, Auf dem H\"ugel 69, 53121 Bonn, Germany}
\address[a2]{Observatoire de Paris-Meudon, 5 place Jules Janssen, 92195 Meudon Cedex, France}
\address[a3]{Harvard-Smithsonian Center for Astrophysics, 60 Garden Street,Cambridge,
Massachusetts 02138, USA}
\address[a4]{Institut f\"ur Theoretische Astrophysik der Universit\"at Heidelberg,
Tiergartenstr. 15, 69121 Heidelberg, Germany, \\
Chatterton Department of Astronomy, School of Physics, University of Sydney
NSW 2006, Australia}
\address[a5]{Sternberg Astronomical Institute, Universitetskii pr. 13,
119899 Moscow, Russia}
\address[a6]{ESO Garching, Karl-Schwarzschild-Str. 2, 85748 Garching bei M\"unchen, Germany.}
%
%
%
\begin{abstract}
We present K$'$-band observations of five Mira stars with the IOTA interferometer.
The interferograms were obtained with the FLUOR fiber optics beam combiner, which
provides high-accuracy
visibility measurements in spite of time-variable atmospheric conditions.
For the
M-type Miras \astrobj{X Oph}, \astrobj{R Aql}, \astrobj{RU Her}, \astrobj{R Ser}, and
the C-type Mira \astrobj{V CrB} we derived the
uniform-disk diameters 11.7~mas, 10.9~mas, 8.4~mas, 8.1~mas, and 7.9~mas
($\pm$0.3 mas), respectively.
Simultaneous photo\-metric observations
yielded the bolometric fluxes. The derived angular Rosseland radii
and the bolometric fluxes allowed the determination of effective
temperatures. For instance, the effective temperature of \astrobj{R Aql} was determined
to be 2970\,$\pm$110~K. A linear Rosseland radius for
\astrobj{R Aql} of 250$^{+100}_{-60}$\,R$_{\odot}$
was derived from the angular Rosseland radius of 5.5\,mas\,$\pm$0.2\,mas and the
HIPPARCOS parallax of 4.73 mas\,$\pm$1.19\,mas.
The observations were compared with
theoretical Mira star models of \citet{bsw96} and \citet{hsw98}.
The effective temperatures of the M-type Miras and the linear radius of \astrobj{R Aql}
indicate fundamental mode pulsation.
\end{abstract}
\end{frontmatter}
%

\section{Introduction}
The resolution of large optical telescopes and interferometers is high enough
to resolve the stellar disk of nearby M giants, to reveal photospheric asymmetries
and surface structures, and to study the dependence of the diameter on
wavelength and variability phase
\citep[see e.g. pioneering work by][]{bon73,karov91}.
Theoretical studies
\citep[e.g.][]{wata79,schol85,bess89,bsw96,hsw98}
show that
accurate monochromatic diameter measurements can significantly improve our understanding
of M giant atmospheres.
With the IOTA (= Infrared-Optical Telescope Array) interferometer,
a resolution of $\sim$\,11.9~mas can be achieved with its largest
baseline of 38~m in the K$'$-band.
The IOTA interferometer is located at the Smithsonian Institution's Whipple Observatory on Mount Hopkins
in Arizona. A detailed description of IOTA can be found in
\citet{carle94} and \citet{trau98}.
IOTA can be operated in the K-band with the
FLUOR (= Fiber Linked Unit for Optical Recombination, \citeauthor{fore97} \citeyear{fore97}) fiber optics beam combiner.
This beam combiner provides high-accuracy visibility measurements
in spite of time-variable atmospheric conditions. The single-mode fibers in the beam combiner
spatially filter the wavefronts corrugated by atmospheric turbulence \citep{fore97,perr98}.

\section{Observations}
\begin{figure*}
\epsfxsize1.0\textwidth
\epsffile{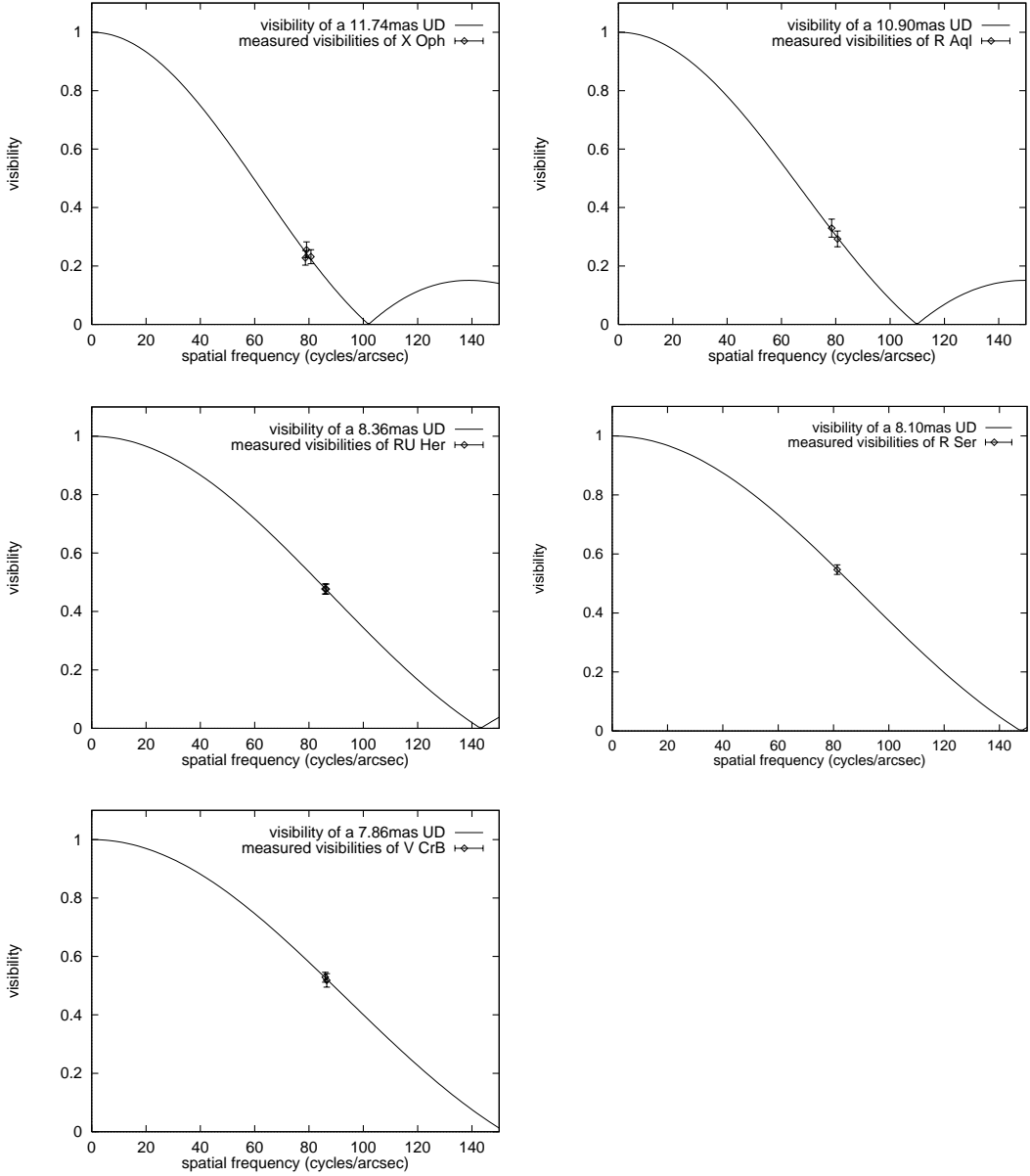}
\caption[]{ Uniform-disk (UD)
fits (X~Oph, R~Aql, RU~Her, R~Ser, and V~CrB).}
\label{fig:Vis}
\end{figure*}

The five Miras \astrobj{X Oph}, \astrobj{R Aql}, \astrobj{RU Her}, \astrobj{R Ser}, \astrobj{V CrB} were observed with the IOTA
interferometer on May 16, 17 and 18, 1999. The observations
were carried out with the fiber optics beam combiner FLUOR
in the K$'$-band (interference filter with center wavelength\,/\,bandwidth of 2.13$\mu$m/0.30$\mu$m) and with
38~m baseline (maximum baseline of the IOTA interferometer). 
The interferograms were scanned by the delay line during the coherence time of the atmosphere.
The OPD length of the scan was $\sim$\,100\,$\mu$m (OPD = Optical Path Difference).
Approximately 100 scans per baseline were recorded.
%
\begin{table} [ht]
\caption[]{Observed data of \astrobj{X Oph}, \astrobj{R Aql}, \astrobj{RU Her}, \astrobj{R Ser} and \astrobj{V CrB}.}
\label{tab:obs}
\begin{center}
{\scriptsize
\begin{tabular}{cccccccccc}
\hline
\rule[-1ex]{0pt}{3.5ex} Star & spectral & $P$ & Date & $\Phi_{\rm vis}$ & $B_{\rm p}$
       & $V$ & $\Theta_{\rm UD}$ & ref. stars & $\Theta_{\rm UD,ref}$ \\
\rule[-1ex]{0pt}{3.5ex}      & type     &  [days] &  &   &  [m] &  & [mas] & [HIP] & [mas] \\
\hline
\rule[-1ex]{0pt}{3.5ex} X Oph & M5e-M9e & 328 & 99 May 17 & 0.71 & 35.47 & 0.2317$\pm$0.024 & 11.74$\pm$0.30 & 86742 & 4.6$\pm$0.5 \\
\rule[-1ex]{0pt}{3.5ex}       &         &     & 99 May 18 &      & 34.75 & 0.2554$\pm$0.027 &  & 98337 & 6.2$\pm$0.6 \\
\rule[-1ex]{0pt}{3.5ex}       &         &     & 99 May 18 &      & 34.57 & 0.2279$\pm$0.025 &  & 98438 & 6.1$\pm$0.6 \\
\rule[-1ex]{0pt}{3.5ex}       &         &     &           &      &       &                  &  & 97278 & 6.8$\pm$0.7 \\
\rule[-1ex]{0pt}{3.5ex} R Aql & M5e-M9e & 284 & 99 May 17 & 0.17 & 35.42 & 0.2927$\pm$0.027 & 10.90$\pm$0.33 & 86742 & 4.6$\pm$0.5 \\
\rule[-1ex]{0pt}{3.5ex}       &         &     & 99 May 18 &      & 34.48 & 0.3295$\pm$0.031 &  & 98337 & 6.2$\pm$0.6 \\
\rule[-1ex]{0pt}{3.5ex}       &         &     &           &      &       &                  &  & 98438 & 6.1$\pm$0.6 \\
\rule[-1ex]{0pt}{3.5ex}       &         &     &           &      &       &                  &  & 97278 & 6.8$\pm$0.7 \\
\rule[-1ex]{0pt}{3.5ex} RU Her & M6e-M9 & 484 & 99 May 17 & 0.07 & 37.95 & 0.4768$\pm$0.017 & 8.36$\pm$0.20 & 71053 & 4.0$\pm$0.4 \\
\rule[-1ex]{0pt}{3.5ex}        &        &     &           &      & 37.73 & 0.4769$\pm$0.017 & & 78159 & 2.8$\pm$0.3 \\
\rule[-1ex]{0pt}{3.5ex} R Ser & M5e-M9e & 356 & 99 May 18 & 0.28 & 35.74 & 0.5467$\pm$0.016 & 8.10$\pm$0.20 & 61658 & 3.7$\pm$0.4 \\
\rule[-1ex]{0pt}{3.5ex}       &            &     &           &      &       &                  &  & 75530 & 3.5$\pm$0.4 \\
\rule[-1ex]{0pt}{3.5ex}       &            &     &           &      &       &                  &  & 85934 & 3.8$\pm$0.4 \\
\rule[-1ex]{0pt}{3.5ex} V CrB & C6,2e(N2e) & 357 & 99 May 16 & 0.07 & 37.78 & 0.5288$\pm$0.017 & 7.86$\pm$0.24 & 73555 & 2.5$\pm$0.3 \\
\rule[-1ex]{0pt}{3.5ex}       &           &      &           &      & 38.02 & 0.5180$\pm$0.023 &  & 81833 & 2.5$\pm$0.3 \\
\hline
\end{tabular}
\rm}
\end{center}
\end{table}
Several reference stars (Table~\ref{tab:obs}) were observed for the calibration of the observations.
The calibrated visibilities of the five Miras were obtained with the FLUOR data reduction
software package described in \citet{fore97} and \citet{perr98}.
Preliminary results of these observations have been
presented in \citet{hof00b}.
Fig.~\ref{fig:Vis}
shows the obtained visibility functions of the five Mira stars together with uniform-disk fits.
The errors of the derived
Mira star diameters are 2.5-3\%.
In Table~\ref{tab:obs} the calibrated visibilities and the derived uniform-disk
diameters of the five Miras are listed, together with observational
parameters (spectral type, variability period $P$, date of observation, variability phase $\Phi_{\rm vis}$
in visual light,
projected baseline length $B_{\rm p}$, calibrated visibilities $V$,
derived uniform-disk diameters $\Theta_{\rm UD}$,
the HIPPARCOS numbers of the reference stars and their uniform disk diameters $\Theta_{\rm UD,ref}$).
All of the observed Mira stars have UD diameters smaller than the
diffraction-limited resolution $\lambda$/b = 11.9~mas of
an interferometer with projected baseline b = 38~m (= largest projected IOTA baseline;
$\lambda$ = 2.13\,$\mu$m).
Therefore, all measured visibility points
lie at spatial frequencies below the first zero of the UD visibility function
of the observed stars
(see Fig.~\ref{fig:Vis}; a baseline length of 38~m and wavelength of 2.13\,$\mu$m
correspond to a spatial
frequency of 85.6 cycles/arcsec). Nevertheless, the UD diameter of the observed stars
can be derived from the fit of a model UD visibility function to the measured visibilities.

\subsection{Calibrated visibilities, stellar diameters, errors}
The transfer function of the whole instrument at the time of the observation
of the object was derived from visibility measurements of several reference stars
(2 to 5 different reference stars per object) before and after the object observations.
Indirect estimates of the diameter of reference stars had to be used, since
only few stellar diameters have been measured up to now.
The diameters of the reference
stars were derived from the list of angular diameters for giants at K=0
according to \citet{dyck96}.
The accuracy of the estimated diameters of
the reference stars is 10\%. The transfer function was directly computed as the
ratio between the measured fringe contrast and the visibility of the reference
star.
The visibility of the reference star was derived from its uniform-disk diameter
estimated by the method of \citet{dyck96}.
Note, that the errors of the target star diameters are much smaller than 10\%.
The error $\sigma_{\Theta_{\rm Obj}}$ of the fitted UD diameter of the target star caused
by the uncertainty
$\sigma_{\Theta_{\rm Ref}}$ of the
diameter of the reference star is given, according to the Gaussian error propagation law, by
\begin{equation}
\frac{\sigma_{\Theta_{\rm Obj}}}{\Theta_{\rm Obj}} = \frac{|V(\Theta_{\rm Obj})|}{|V(\Theta_{\rm Ref})|} \cdot \
\frac{|\partial V(\Theta_{\rm Ref})/\partial \Theta|}{|\partial V(\Theta_{\rm Obj})/\partial \Theta|} \cdot \
\frac{\Theta_{\rm Ref}}{\Theta_{\rm Obj}} \cdot \frac{\sigma_{\Theta_{\rm Ref}}} {\Theta_{\rm Ref}},
\end{equation}
where $V(\Theta_{\rm Obj})$ and $V(\Theta_{\rm Ref})$ are the visibilities
of the target and reference star, respectively.
If, for example, one considers \astrobj{R Ser}, then $\Theta_{\rm Obj}$=8.10\,mas, $\Theta_{\rm Ref}$=3.81\,mas,
$V(\Theta_{\rm Obj})$=0.55, $V(\Theta_{\rm Ref})$=0.88,
$\partial V(\Theta_{\rm Obj})/\partial \Theta$=0.09\,mas$^{-1}$,
and $\partial V(\Theta_{\rm Ref})/\partial \Theta$=0.06\,mas$^{-1}$. According to Eq.(1),
the relative error of the UD diameter of the (larger) target star is 2\%,
while the diameter error of the reference star is 10\%.

The total errors of the calibrated visibilities presented in Table~\ref{tab:obs} take into
account the following errors:
\begin{itemize}
\item the statistical error of the raw visibility measurement in each star (e.g. photon and detector noise);
\item the errors of the measured gain matrix (link between photometric and
interferometric signals, \citeauthor{fore97} \citeyear{fore97});
\item the uncertainty on the predicted visibility of the reference star
  (the error of the estimated diameter of the reference star was assumed
   to be 10\% according to \citeauthor{dyck96} \citeyear{dyck96});
\item the uncertainties of the transfer function at the time of the object measurement
  (the transfer function is interpolated between measurements of the
  reference star before and after the object observation; see details in \citeauthor{perr98}
\citeyear{perr98}).
\end{itemize}
In addition to the reduction, based on all available reference stars,
each target star was also separately reduced with each of the 2-5 different reference stars.
All these calibrated visibilities are within the error bars given in Table~\ref{tab:obs}.

The uniform-disk diameters and the diameters corresponding to a model center-to-limb
intensity variation (hereafter CLV) were derived from the calibrated visibilities with a
$\chi^2$ fit:
\begin{equation}
\chi^2 = \sum_{\rm i=1}^{\rm N} \frac{|V_{\rm i} - V(\Theta, u_{\rm i})|^2} {\sigma_{\rm i}^2},
\end{equation}
where $V_{\rm i}$ denotes the measured visibility at spatial frequency $u_{\rm i}$
with the error $\sigma_{\rm i}$,
$V(\Theta, u_{\rm i})$ is the visibility of the theoretical CLV (uniform-disk or any model CLV)
at frequency $u_{\rm i}$
as a function of the disk diameter $\Theta$, and N denotes the number of measured visibilities.
The diameter error $\sigma_{\Theta_{\rm Obj}}$ of the fitted stellar diameter $\Theta_{\rm Obj}$
is linked, according to the Gaussian error propagation law,
to the total errors $\sigma_{\rm i}$ of the derived object visibilities $V_{\rm i}$
by
\begin{equation}
1/\sigma_{\Theta_{\rm Obj}}^2 = \sum_{\rm i=1}^{\rm N} |a_{\rm i}|^2 / \sigma_{\rm i}^2,
\end{equation}
where $a_{\rm i}$ denotes the derivative $\partial V(\Theta_{\rm Obj}, u_{\rm i}) / \partial \Theta$
of the visibility of the theo\-retical CLV at the fitted disk diameter $\Theta_{\rm Obj}$.
The diameter errors listed in Table~\ref{tab:obs} and used below are derived
from the above relation.

\section{Comparison of the observations with Mira star models}
In this section we derive
angular diameters
from the measured visibilities by fitting
different theoretical center-to-limb intensity variations
of different Mira star models 
(\citet{bsw96} = BSW96, \citet{hsw98} = HSW98).
From these angular diameters
and the bolometric fluxes, we derive effective temperatures.
For \astrobj{R Aql}
a HIPPARCOS parallax is available, which allows us to
determine linear radii.
The comparison of these measured stellar parameters with theoretical ones,
indicate whether any of the models is a fair representation of the
observed Mira stars.
\begin{table} [ht]
\caption[]{Properties of Mira model series (see text).}
\label{tab:prop}
\begin{center}
{\scriptsize
\begin{tabular}{ccccccc}
\hline
\rule[-1ex]{0pt}{3.5ex} Series & Mode & $P$(d) & $M/M_{\odot}$ & $L/L_{\odot}$
       & $R_{\rm p}/R_{\odot}$ & $T_{\rm eff}$/K \\
\hline
\rule[-1ex]{0pt}{3.5ex} D & f & 330 & 1.0 & 3470 & 236 & 2900  \\
\rule[-1ex]{0pt}{3.5ex} E & o & 328 & 1.0 & 6310 & 366 & 2700  \\
\rule[-1ex]{0pt}{3.5ex} P & f & 332 & 1.0 & 3470 & 241 & 2860  \\
\rule[-1ex]{0pt}{3.5ex} M & f & 332 & 1.2 & 3470 & 260 & 2750  \\
\rule[-1ex]{0pt}{3.5ex} O & o & 320 & 2.0 & 5830 & 503 & 2250  \\
\hline
\end{tabular}
\rm}
\end{center}
\end{table}
\begin{table} [ht]
\caption[]{Link between the 22 abscissa values (model-phase combinations m)
in Figs.~\ref{fig:linradii} and \ref{fig:Teff}, and the models.
The variability phase
$\Phi_{\rm vis}$ in visual light ($\Phi_{\rm vis}$ represents cycle + phase),
the Rosseland radius $R$ and the K$'$-band radius $R_{\rm K'}$
in units
of the parent star radius $R_p$, and the effective temperature $T_{\rm eff} \propto (L/R^2)^{1/4}$
are also given.}
\label{tab:link}
\begin{center}
{\scriptsize
\begin{tabular}{llllll}
\hline
\rule[-1ex]{0pt}{3.5ex} Model & $\Phi_{\rm vis}$ & $R/R_{\rm p}$ & $R_{\rm K'}/R_{\rm p}$ & $T_{\rm eff}$/K & m \\
\hline
\rule[-1ex]{0pt}{3.5ex} D27520 & 1+0.0  &  1.04 & 1.02  & 3020 & 1 \\
\rule[-1ex]{0pt}{3.5ex} D27760 & 1+0.5  &  0.91 & 0.90  & 2710 & 2 \\
\rule[-1ex]{0pt}{3.5ex} D28760 & 2+0.0  &  1.04 & 1.02  & 3030 & 3 \\
\rule[-1ex]{0pt}{3.5ex} D28960 & 2+0.5  &  0.91 & 0.91  & 2690 & 4 \\
\rule[-1ex]{0pt}{3.5ex} E8300  & 0+0.83 &  1.16 & 1.14  & 2330 & 5 \\
\rule[-1ex]{0pt}{3.5ex} E8380  & 1+0.0  &  1.09 & 1.11  & 2620 & 6 \\
\rule[-1ex]{0pt}{3.5ex} E8560  & 1+0.21 &  1.17 & 1.14  & 2610 & 7 \\
\rule[-1ex]{0pt}{3.5ex} P71800 & 0+0.5  &  1.20 & 1.03  & 2160 & 8 \\
\rule[-1ex]{0pt}{3.5ex} P73200 & 1+0.0  &  1.03 & 0.98  & 3130 & 9 \\
\rule[-1ex]{0pt}{3.5ex} P73600 & 1+0.5  &  1.49 & 1.08  & 1930 & 10 \\
\rule[-1ex]{0pt}{3.5ex} P74200 & 2+0.0  &  1.04 & 1.12  & 3060 & 11 \\
\rule[-1ex]{0pt}{3.5ex} P74600 & 2+0.5  &  1.17 & 1.01  & 2200 & 12 \\
\rule[-1ex]{0pt}{3.5ex} P75800 & 3+0.0  &  1.13 & 1.05  & 3060 & 13 \\
\rule[-1ex]{0pt}{3.5ex} P76200 & 3+0.5  &  1.13 & 0.95  & 2270 & 14 \\
\rule[-1ex]{0pt}{3.5ex} P77000 & 4+0.0  &  1.17 & 1.14  & 2870 & 15 \\
\rule[-1ex]{0pt}{3.5ex} M96400 & 0+0.5  &  0.93 & 0.92  & 2310 & 16 \\
\rule[-1ex]{0pt}{3.5ex} M97600 & 1+0.0  &  1.19 & 1.15  & 2750 & 17 \\
\rule[-1ex]{0pt}{3.5ex} M97800 & 1+0.5  &  0.88 & 0.91  & 2460 & 18 \\
\rule[-1ex]{0pt}{3.5ex} M98800 & 2+0.0  &  1.23 & 1.19  & 2650 & 19 \\
\rule[-1ex]{0pt}{3.5ex} O64210 & 0+0.5  &  1.12 & 1.08  & 2050 & 20 \\
\rule[-1ex]{0pt}{3.5ex} O64530 & 0+0.8  &  0.93 & 0.95  & 2150 & 21 \\
\rule[-1ex]{0pt}{3.5ex} O64700 & 1+0.0  &  1.05 & 1.01  & 2310 & 22 \\
\hline
\end{tabular}
\rm}
\end{center}
\end{table} 
All Mira star models used in this paper are from BSW96 (D and E series)
and from HSW98 (P, M and O series). They were developed as possible representations
of the prototype Mira variable o~Ceti, and hence have periods $P$
very close to the 332 day period of this star; they differ in pulsation mode, assumed mass $M$ and
assumed luminosity $L$; and the BSW96 models differ from the (more advanced) HSW98 models
with respect to the pulsation modelling technique.
The five models represent stars pulsating in the fundamental mode ($f$; D, P and M models) or
in the first-overtone mode ($o$; E and O models).
Table~\ref{tab:prop} lists the properties of these Mira model series ($R_{\rm p}$ =
Rosseland radius of the
non-pulsating "parent" star of the Mira variable  [see BSW96/HSW98] , i.e. the
distance from the parent star's center, at which the Rosseland optical depth $\tau_{\rm Ross}$ equals
unity;
$T_{\rm eff} \propto (L/R_{\rm p}^2)^{1/4}$ = effective
temperature; L = luminosity).
Table \ref{tab:link} provides the link between the 22 abscissa values
(model-phase combinations m)
in Figs.~\ref{fig:linradii} and \ref{fig:Teff},
and the models, and it additionally lists the variability phase, Rosseland
and stellar K$'$-band filter radius (definition given below)
in units of the parent star radius $R_{\rm p}$,
and the effective temperature.
We compare predictions of these models at different phases and cycles with
our measurements.

\subsection{Monochromatic radius $R_{\lambda}$,
Rosseland radius $R$ and stellar filter radius $R_{\rm f}$}
{\it Monochromatic radius $R_{\lambda}$}: We use the conventional stellar radius definition where
the monochromatic radius $R_{\lambda}$ of a star
at wavelength $\lambda$ is given by the distance from the star's
center at which the optical depth equals unity ($\tau_{\lambda}$\,=\,1).

{\it Rosseland radius $R$}: In analogy, the photospheric stellar radius or
Rosseland radius $R$ is given by the
distance from the star's center at which the Rosseland optical depth
equals unity ($\tau_{\rm Ross}$\,=\,1).

{\it Stellar filter radius $R_{\rm f}$ and $R_{\rm K'}$}: The stellar filter radius $R_{\rm f}$
for filter transmission ${\rm f}_{\lambda}$
is the intensity and filter weighted radius
$R_{\rm f} = \int R_{\lambda}\,I_{\lambda}\,{\rm f}_{\lambda}\,d\lambda\,/\,\int I_{\lambda}\,{\rm f}_{\lambda}\,d\lambda$,
which we call stellar filter radius $R_{\rm f}$
after the definition of \citet{schol87}.
In this equation
$R_{\lambda}$ denotes the monochromatic $\tau_{\lambda}$\,=\,1 radius,
$I_{\lambda}$ the central intensity spectrum and ${\rm f}_{\lambda}$ the transmission of the filter.
We have calculated the theoretical CLVs of the above-mentioned Mira star models (D, E, P, M and O)
at different phases and cycles for the filter K$'$
(center wavelength\,/\,bandwidth of 2.13\,$\mu$m\,/\,0.30\,$\mu$m) used for the observations.
In the near-continuum K$'$-band
window, within which the extinction coefficient varies very
little with wavelength in most models, monochromatic radii $R_{\lambda}$ vary very
little and are almost identical to the K$'$-band filter radius $R_{\rm K'}$.

Note that the
Rosseland radius $R$ is a non-observable quantity based on the Rosseland mean of
extinction coefficients, and its correlation with monochromatic ($R_{\lambda}$) or filter radii
(e.g. $R_{\rm K'}$) has to be deduced from a model. Often, the K$'$-band radius and other
near-continuum filter radii are close to the Rosseland radius (e.g. HSW98).
Some of the models, however, predict wavelength-dependent CLV shapes
within the K$'$-band due to molecular (mainly water) absorption, resulting in a
two-component structure (i.e. disk plus tail) of the CLV \citep{schol00,bed01}
and contributing to a few noticeable differences between $R$ and 
$R_{\rm K'}$, seen in Table~\ref{tab:link}.

\subsection{Derived angular stellar K$'$-band radius $R_{\rm K',m}^a$ (a\,=\,angular,
K$'$\,=\,K$'$-band, m\,=\,model-phase combination)
and angular Rosseland radius $R_{\rm m}^a$ of the observed Miras}
The {\it derived angular stellar K$'$-band radii $R_{\rm K',m}^a$}
(corresponding to the
model-phase combinations m; Table~\ref{tab:link}) were determined by
least-squares fits between the
measured Mira star visibilities (Fig. 1) and the visibilities of the corresponding model CLVs
(of the models described in Table 3).

Additionally, the {\it angular Rosseland radii $R_{\rm m}^a$}
were derived from the obtained
stellar K$'$-band radii $R_{\rm K',m}^a$ and the theoretical ratios
between $R$/$R_{\rm p}$ and $R_{\rm K'}$/$R_{\rm p}$ given in Table~\ref{tab:link}.
In the following subsections, we apply CLVs predicted from all five models at phases both
near our observations and, for comparison, also at other phases.
\subsection{Linear radii}
\begin{figure*}
\epsfxsize1.0\textwidth
\epsffile{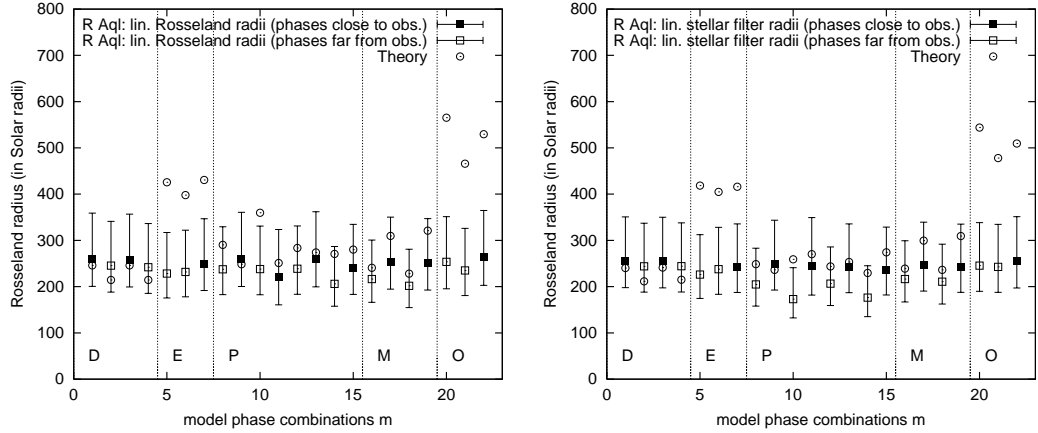}
\caption[]{ Comparison of measured \astrobj{R Aql} radii and theoretical model radii:
(left) linear Rosseland radii $R_{\rm m}$ and (right) linear stellar K$'$-band radii $R_{\rm K',m
}$
for all 22 model-phase combinations m.
Measured linear radii derived from models {\it with phases close to our observations (= filled squares)}
and
far from our observations (open squares) are shown. The theoretical model radii are plotted with
open circles.
Table \ref{tab:link} gives the link between the abscissa values (model-phase combinations m)
and the models and their phases. }
\label{fig:linradii}
\end{figure*}

We have derived linear stellar K$'$-band radii $R_{\rm K',m}$ and linear Rosseland
radii $R_{\rm m}$ of \astrobj{R Aql} from the measured angular stellar K$'$-band radii
$R_{\rm K',m}^a$
and Rosseland radii $R_{\rm m}^a$ by using the \astrobj{R Aql} HIPPARCOS parallax
of 4.73$\pm$1.19\,mas \citep{ESA97,leeuw97}.
The HIPPARCOS parallaxes of the
other four observed Miras have too large errors
to reliably estimate stellar linear radii.
Fig.~\ref{fig:linradii}
shows the obtained linear Rosseland radii $R_{\rm m}$ and stellar K$'$-band radii $R_{\rm K',m}$
of \astrobj{R Aql} for all model-phase combinations m.
The theo\-retical Rosseland radii of the D, M
and P fundamental mode
model series at all available near-maximum phases are close
(i.e. within the error bars)
to the measured Rosseland radii of \astrobj{R Aql}.
The theoretical Rosseland radii of the first-overtone models E and O are clearly
too large compared with
measured Rosseland radii.
The same conclusions are also valid for the linear stellar filter radii
$R_{\rm K'}$ (Fig.~\ref{fig:linradii}).

Table~\ref{tab:Linear} lists measured linear Rosseland radii of \astrobj{R Aql}
derived from the well-fitting near-maximum
fundamental mode models D, P and M and the corresponding theoretical ones.
The model-averaged measured linear Rosseland radii are averages over those
cycles of the model where the measured and theoretical linear Rosseland radius values are
within the error bar (see Fig. ~\ref{fig:linradii}).
\begin{table}
\caption[]{
Measured linear Rosseland radii of \astrobj{R Aql} for the well-fitting near-maximum
fundamental mode models D, P and M
and a comparison with theory.
The model-averaged measured linear Rosseland radii are averages over those
cycles (at phases close to our observation) of the model where the measured and theoretical linear
Rosseland radius values are
within the error bar.
For comparison, the linear uniform disk radius of
\astrobj{R Aql} is 248$_{-56}^{+93}$\,R$_{\odot}$.
}
\label{tab:Linear}
\begin{center}
{\scriptsize
\begin{tabular}{ccc}
\hline
\rule[-1ex]{0pt}{3.5ex} model & measured linear & theoretical linear \\ 
\rule[-1ex]{0pt}{3.5ex}       & Rosseland radius (R$_{\odot}$)   & Rosseland radius (R$_{\odot}$) \\
\hline
\rule[-1ex]{0pt}{3.5ex} D & 259$^{+99}_{-59}$ & 246 \\
\rule[-1ex]{0pt}{3.5ex} P & 246$^{+100}_{-60}$ & 263 \\
\rule[-1ex]{0pt}{3.5ex} M & 252$^{+97}_{-58}$ & 315 \\
\hline
\end{tabular}
\rm}
\end{center}
\end{table}
\begin{figure*}
\epsfxsize1.0\textwidth
\epsffile{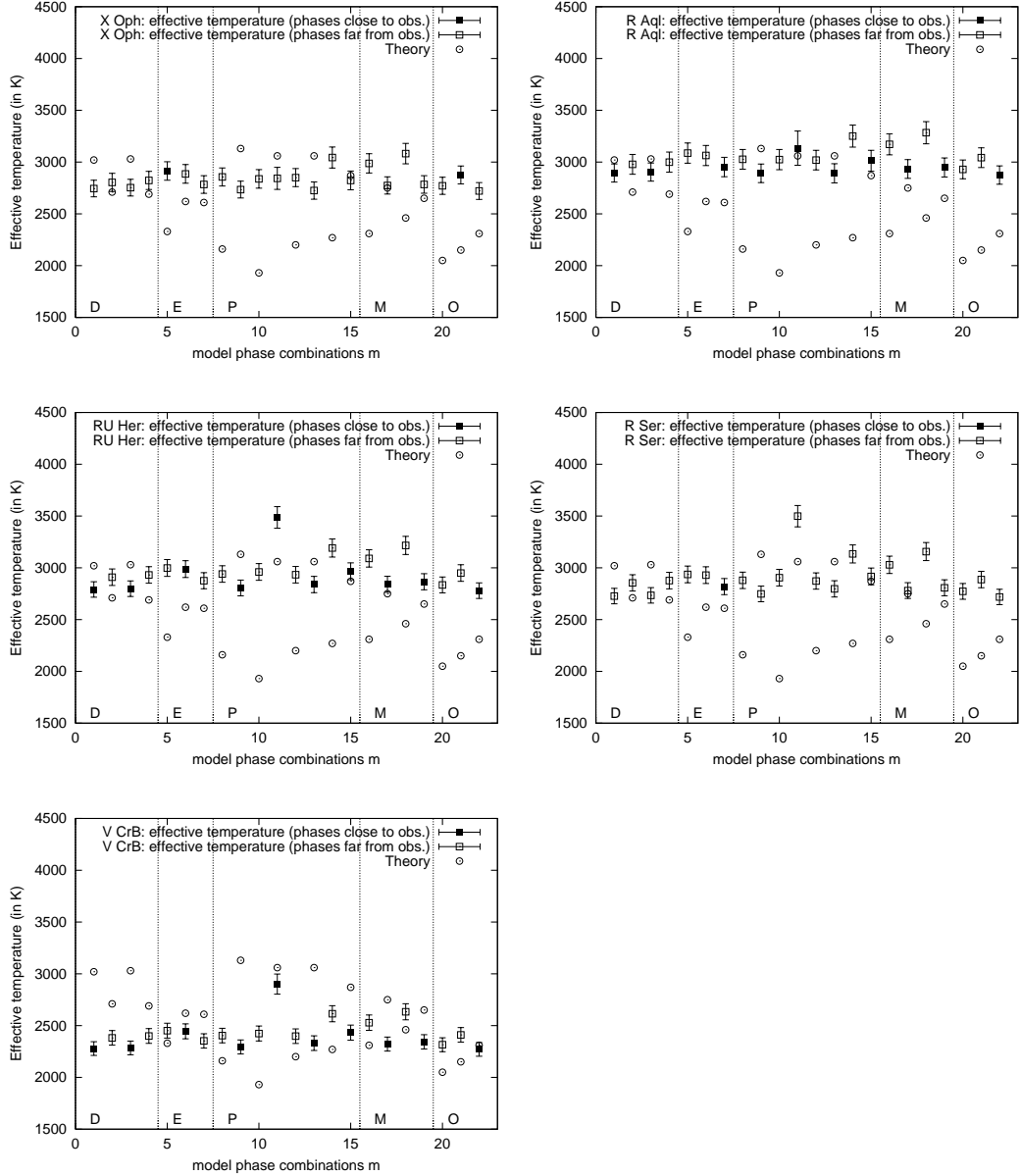}
\caption[]{
Comparison of measured effective temperatures
of the 5 observed Mira stars and the model effective temperatures (see text).
The measured effective temperatures are derived from the angular Rosseland radii determined
by least-square fits between the measured visibilities and the visibilities of the corresponding
theoretical CLVs given by the models of BSW96 \& HSW98.
The measured T$_{\rm eff}$ values are given with 1$\sigma$-error bars.
Measured T$_{\rm eff}$ values derived from models {\it at phases close to our observations
(= filled squares)}
and far from our observations (open squares) are shown. The theoretical model effective
temperatures are plotted with open circles.
Table \ref{tab:link} shows the link between the abscissa values and the models and their
phases.
}
\label{fig:Teff}
\end{figure*}
\subsection{Effective temperature}
\begin{table}
\caption[]{Observational data and measured effective temperatures. For each star
the effective temperature according to its UD diameter is listed. For \astrobj{R Aql} and
\astrobj{RU Her} measured $T_{\rm eff}$ values are given which were derived from the Rosseland
radii of those model-phase
combinations (at phases close to our observation), where the theoretical $T_{\rm eff}$ value
and the measured one are
within the 2$\sigma$-error bar (see Fig. 3).
The indices D, P, M refer to the pulsation models.
}
\label{tab:Fboltab}
\begin{center}
{\scriptsize
\begin{tabular}{cccccllll}
\hline
\rule[-1ex]{0pt}{3.5ex} Star & Date & $\Phi_{\rm vis}$ & $K$   & F$_{\rm bol}$              & $T_{\rm eff,UD}$ & $T_{\rm eff,D}$& $T_{\rm eff,P}$ & $T_{\rm eff,M}$ \\
\rule[-1ex]{0pt}{3.5ex}      & 1999 &                  & [mag] & [$10^{-8}$\,erg/cm$^2$\,s] & [K]              & [K]            & [K]             & [K] \\
\hline
\rule[-1ex]{0pt}{3.5ex} X Oph & May 27 & 0.71 & -0.83 & 287.7$\pm$28.8 & 2810$\pm$90 & - & - & - \\
\rule[-1ex]{0pt}{3.5ex} R Aql & May 28 & 0.17 & -0.86 & 305.6$\pm$30.6 & 2960$\pm$100 & 2900$\pm$90 & 3010$\pm$120 & - \\
\rule[-1ex]{0pt}{3.5ex} RU Her & May 21 & 0.07 & -0.11 & 153.2$\pm$15.3 & 2850$\pm$80 & - & 2970$\pm$80 & 2840$\pm$80 \\
\rule[-1ex]{0pt}{3.5ex} R Ser & May 21 & 0.28 & 0.02 &  130.3$\pm$13.1 & 2780$\pm$80 & - & - & - \\
\rule[-1ex]{0pt}{3.5ex} V CrB & May 27 &  0.07 & 0.96 &   59.9$\pm$6.0 & 2320$\pm$70 & - & - & - \\
\hline
\end{tabular}
\rm}
\end{center}
\end{table} 
Effective temperatures of each observed Mira star were derived from its angular
Rosseland radii $R_{\rm m}^a$
and
its bolometric flux using the relation
\begin{equation}
T_{\rm eff} = 2341~{\rm K} \times (F_{\rm bol}/\phi^2)^{1/4}
\end{equation}
where $F_{\rm bol}$ is the apparent bolometric flux in units of 10$^{-8}$\,erg\,cm$^{-2}$\,s$^{-1}$
and $\phi$\,=\,2$\times$$R_{\rm m}^a$ is the angular Rosseland diameter in mas.
The bolometric fluxes were derived from optical ({\it VR}), near-infrared ({\it JHKLM}) and
mid-infrared (e.g. IRAS data) photometry (see Appendix A).
Interstellar extinction corrections were applied which, however, affect the fluxes only mildly
(see Table~\ref{tab:Tphoto}).
The near-infrared photometry was carried out with the 1.25~m telescope at the Crimean station of
the Sternberg Astronomical Institute in Moscow 12 days after our visibility observations.
The V band photometry was taken from the AAVSO data base \citep{matt00}
at the time of our observations.
The R band data were derived from observations with the 1.25~m telescope at the Crimean station of
the Sternberg Astronomical Institute and the AAVSO data base.
The conventional approximation for calculating bolometric fluxes is to use a black body function
to interpolate between photometric observations in the near-infrared ({\it JHKLM}).
However, applying only this black body as a description of the whole spectral energy distribution,
yields bolometric fluxes being in most cases significantly larger
than those based also on additional photometry in $V$ and mid-infrared.
In Appendix A, the procedure for deriving bolometric fluxes used in this paper is described in detail.
The errors of the {\it JHKLM} photometry do not exceed 0.03 magnitudes, whereas 
the errors of the V and R photometry are approx. 0.6 magnitudes, which affect the bolometric fluxes.
The errors of the bolometric fluxes were determined by integrating the SED derived
from the photometric data
at the upper and lower boundary of the error bar. The derived bolometric fluxes
have errors of approx. 10\%.

Fig.~\ref{fig:Teff} shows a comparison of the measured and theoretical effective temperatures.

\noindent
(a) For the two M stars \astrobj{X Oph} and \astrobj{R Ser}, the phases of the models do not fit
the phases at observations
(\astrobj{X Oph}: only the E model at phase 0.83 and the O model at phase 0.80 
are close to the phase 0.71
at observation; \astrobj{R Ser}: only the E model at phase 1.21 
is close to the phase 0.28 at observation).

\noindent
(b) For the C-type Mira \astrobj{V CrB}, the information drawn from Fig. 3 should be
considered with due caution, because the K$'$-band continuum is slightly 
contaminated by molecular lines in both M and C stars, and because even the 
pure-continuum CLV might be influenced by the different structures of M- and 
C-type atmospheres. The good description by the high-mass, first-overtone O 
model may be a chance match.

\noindent
(c) For the remaining stars, \astrobj{R Aql} and \astrobj{RU Her},
all five models are available at phases close
to the observations (phase 0.17 for \astrobj{R Aql} and 0.07 for \astrobj{RU Her}).
Note, however, that \astrobj{RU Her} has an appreciably longer period than the 
$o$~Ceti period adopted for all model series. For \astrobj{R Aql} and \astrobj{RU Her},
phase- and model-dependent effective temperatures could be derived, which are
listed in Table~\ref{tab:Fboltab}.

\noindent
(d) For the smaller stars \astrobj{RU Her} and \astrobj{R Ser}, extremely
large measured effective temperatures are seen in Fig. 3 at 
model-phase combination m=11 (near-maximum model P74200).
The P74200 model predicts a very pronounced two-component, disk-plus-tail CLV 
structure, as mentioned above, which essentially distorts the central maximum 
of the visibility of the tail-free curve at smaller spatial frequencies. For 
the small disks of \astrobj{RU Her} and \astrobj{R Ser}, the large measured visibilities (V = 0.48 
and 0.55, respectively) lie in the inner portion of the central maximum, which
should be mostly affected by a two-component CLV structure. Obviously, such a
pronounced structure is not present in these stars. For the larger disks of
\astrobj{X Oph} and \astrobj{R Aql}, the smaller values of V = 0.23 and 0.29, respectively, belong
to higher spatial frequencies at which the influence of a CLV tail, if it 
were present, would not be very conspicuous. Observations with several baselines
would be necessary for detecting tail-generated distortions of the visibility
curve.

\noindent
(e) Interestingly, the visibility fit of \astrobj{V CrB} whose disk size is similar to
those of \astrobj{RU Her} and \astrobj{R Ser} would be compatible with a distinct two-component 
brightness distribution, but no model studies of limb-darkening of C-type Miras
in the K (K$'$) bandpass are available at the present.

The four M stars \astrobj{X Oph}, \astrobj{R Aql}, \astrobj{RU Her} and \astrobj{R Ser}
clearly do not fit the first-overtone models E and O.
For the two M stars, \astrobj{R Aql} and \astrobj{RU Her}, observed at near-maximum phase, however,
the measured and theoretical T$_{\rm eff}$ values are within
the 2$\sigma$-error bar for some cycles of the fundamental mode models D, P and M
(\astrobj{R Aql}: all cycles of the D model and 3 of 4 cycles of the P model;
\astrobj{RU Her}: 1 of 4 cycles of the P model and 1 of 2 cycles of the M model).
Table~\ref{tab:Fboltab} lists the measured bolometric flux, the T$_{\rm eff}$ values
derived from the measured UD diameter, and for the two M stars \astrobj{R Aql} and \astrobj{RU Her}
measured T$_{\rm eff}$ values for
selected models close to the phase at observation. These
model-averaged T$_{\rm eff}$ values were obtained by averaging only those values
where the theoretical T$_{\rm eff}$ values and the measured ones are within the 2$\sigma$-error bar.
The T$_{\rm eff}$ values derived from the UD diameters are in good agreement 
with the model-derived effective temperatures.

\section{Discussion}
We derived angular uniform-disk diameters $\Theta_{\rm UD}$
of five Mira stars (Table~\ref{tab:obs}) from
K$'$-band visibility measurements with the IOTA interferometer and the FLUOR beam combiner
at 38\,m baseline.
Using simultaneously observed bolometric fluxes and the measured uniform-disk diameters 
we obtained T$_{\rm eff,UD}$ values given in Table~\ref{tab:Fboltab}.

Previous interferometric K-band observations of some of our target stars
(\astrobj{R Aql}, \astrobj{X Oph}, \astrobj{R Ser}) were carried out
by \citet{bell96}
at similar phases.
Their derived uniform-disk diameters (\astrobj{R Aql}; $\Phi_{\rm vis}$\,=\,0.90: 10.76$\pm$0.61\,mas,
\astrobj{X Oph}; $\Phi_{\rm vis}$\,=\,0.75: 12.30$\pm$0.66\,mas,
\astrobj{R Ser}; $\Phi_{\rm vis}$\,=\,0.32: 8.56$\pm$0.58\,mas) are in good agreement
with our observations.
Their effective temperatures, derived from measured angular Rosseland radii
(\astrobj{R Aql}: 3189$\pm$147\,K, \astrobj{X Oph}: 3041$\pm$160\,K,
\astrobj{R Ser}: 2804$\pm$144\,K), are also
in agreement with our results.

The comparison of the observations with Mira star models, with respect to the
effective temperature, suggests
that the four observed M Miras can approximately be represented by the
fundamental mode models D, P or M, whereas the overtone models are much too 
cool.
For the two M stars \astrobj{R Aql} and \astrobj{RU Her},
phase- and model-dependent effective temperatures could be derived, which are
listed in Table~\ref{tab:Fboltab}. These effective temperatures are within the error bars
of the T$_{\rm eff}$ values obtained with the measured uniform disk diameter.
Any more accurate model interpretation of our four 
M-type stars would require an extension of the parameter range (Table~\ref{tab:prop}) and refining
the phase spacing (Table~\ref{tab:link}) of available Mira model grids.
A quantitative study of \astrobj{V CrB} can only be given on the basis of C-type
Mira models which are not yet available.

For \astrobj{R Aql}, a useful HIPPARCOS parallax (4.73$\pm$1.19\,mas) is available and it is therefore possible to
compare measured linear Rosseland and stellar K$'$-band radii with the theoretical radii of the BSW96 and
HSW98 models.
The measured radii were derived by fitting theoretical (BSW96, HSW98)
center-to-limb intensity
variations to the visibility data.
In Table~\ref{tab:Linear}, the measured linear Rosseland radii derived from the
well fitting near-maximum fundamental mode models D, P and M are listed.
From the measured linear Rosseland radii of \astrobj{R Aql} the pulsation mode could not
be determined because of
the large parallax error.

The comparison suggests that \astrobj{R Aql} can well be represented by the 
fundamental mode D or P model.
Note, however, that observations in more filters than just one continuum filter and
more baselines may be necessary
for safely distinguishing a well-fitting model from an accidental match
(cf. Hofmann et al. 2000a). Furthermore, in order to discover the tail structure predicted by some
models, future observations should cover more baselines.
\section{Acknowledgements}
We acknowledge usage of
observations from the AAVSO international database.
This research has made use of the SIMBAD database, operated by CDS in
Strasbourg.
%
\appendix

\section{Bolometric fluxes}
The bolometric fluxes are  determined by integrating
the spectral energy distributions (SEDs).
The present $JHKLM$ photometry were complemented by coeval $V$ data of the
AAVSO \citep{matt00}. Inspection of $BVRJ$ photometry taken
one pulsational cycle later, nearly at the same phase,
shows good agreement with the AAVSO data. From
this data 
the $R$ fluxes at epoch and cycle of the present observations were
interpolated. Table~\ref{tab:Tphoto} presents the $V$ (AAVSO),
$R$ (interpolated) and $JHKLM$ magnitudes.
In the long-wavelength regime the respective IRAS measurements were
taken into account, and  in the case of \astrobj{R Aql} also 8-30\,$\mu$m
photometry of \citet{epch80}
taken at almost the same phase as our data ($\Phi=0.11$).
Interstellar extinction corrections, $A_{\rm V}$, are
given by \citet{whit00} and were considered
by adopting the method of \citet{sav79} with
$A_{\rm V} = 3.1 E(B-V)$. However, extinction is generally 
small (see Table~\ref{tab:Tphoto}) and affects the fluxes only
mildly.
%
\begin{table}[bhtp]
\caption{
Pulsational phase, $\Phi_{\rm vis}$, photometric data in the
$V$ ($0.55\,\mu$m, AAVSO),
$R$ ($0.71\,\mu$m, interpolated),
$J$ ($1.25\,\mu$m),
$H$ ($1.62\,\mu$m),
$K$ ($2.20\,\mu$m),
$L$ ($3.50\,\mu$m), and $M$ ($4.80\,\mu$m) band (in mag), adopted extinction
$A_{V}$ (in mag), total bolometric flux, $F_{\rm bol}$,
based on a three-component fit to the
$VRJ$, $JHKLM$ and mid-infrared data, flux difference
$\delta F_{\rm bol}^{\rm bb}$ of a SED fit based only
on $JHKLM$ photometry, and flux contribution $\Delta F_{\rm bol}^{\rm dust}$
due to dust emission.
}
\label{tab:Tphoto}
\flushleft
{\tiny
\begin{tabular}{lcccccccccccc}
\hline
Star    & $\Phi_{\rm vis}$ & V    &  R   &  J   &    H  &    K  &   L   &    M  & $A_{\rm V}$ & $F_{\rm bol}$
        & $\delta F_{\rm bol}^{\rm bb}$ &  $\Delta F_{\rm bol}^{\rm dust}$  \\
        &                  &      &      &      &       &       &       &       &             & [$10^{-8}$\,erg/cm$^2$\,s]
        &                               &                                   \\ 
\hline
X\,Oph  & 0.71   & 8.30 & 5.43 & 0.55 & -0.32 & -0.83 & -1.49 & -1.25 &  0.18 & 287.7 & +26.2\% & 2.4\% \\
R\,Aql  & 0.17   & 7.90 & 5.01 & 0.47 & -0.40 & -0.86 & -1.40 & -1.32 &  0.23 & 305.6 & +29.5\% & 1.5\% \\
RU\,Her & 0.07   & 7.70 & 5.12 & 1.28 &  0.53 & -0.11 & -0.82 & -0.66 &  0.04 & 153.2 & +12.9\% & 1.8\% \\
R\,Ser  & 0.28   & 9.90 & 6.55 & 1.26 &  0.45 &  0.02 & -0.51 & -0.30 &  0.04 & 130.3 & +40.7\% & 2.7\% \\
V\,CrB  & 0.07   & 8.10 & 5.73 & 2.79 &  1.68 &  0.96 & -0.10 & -0.13 &  0.03 & 59.9  & - 6.0\% & 2.4\% \\
\hline
\end{tabular}
\rm}
\end{table}

\begin{figure}
\centering
\epsfxsize=5cm
\mbox{\epsffile{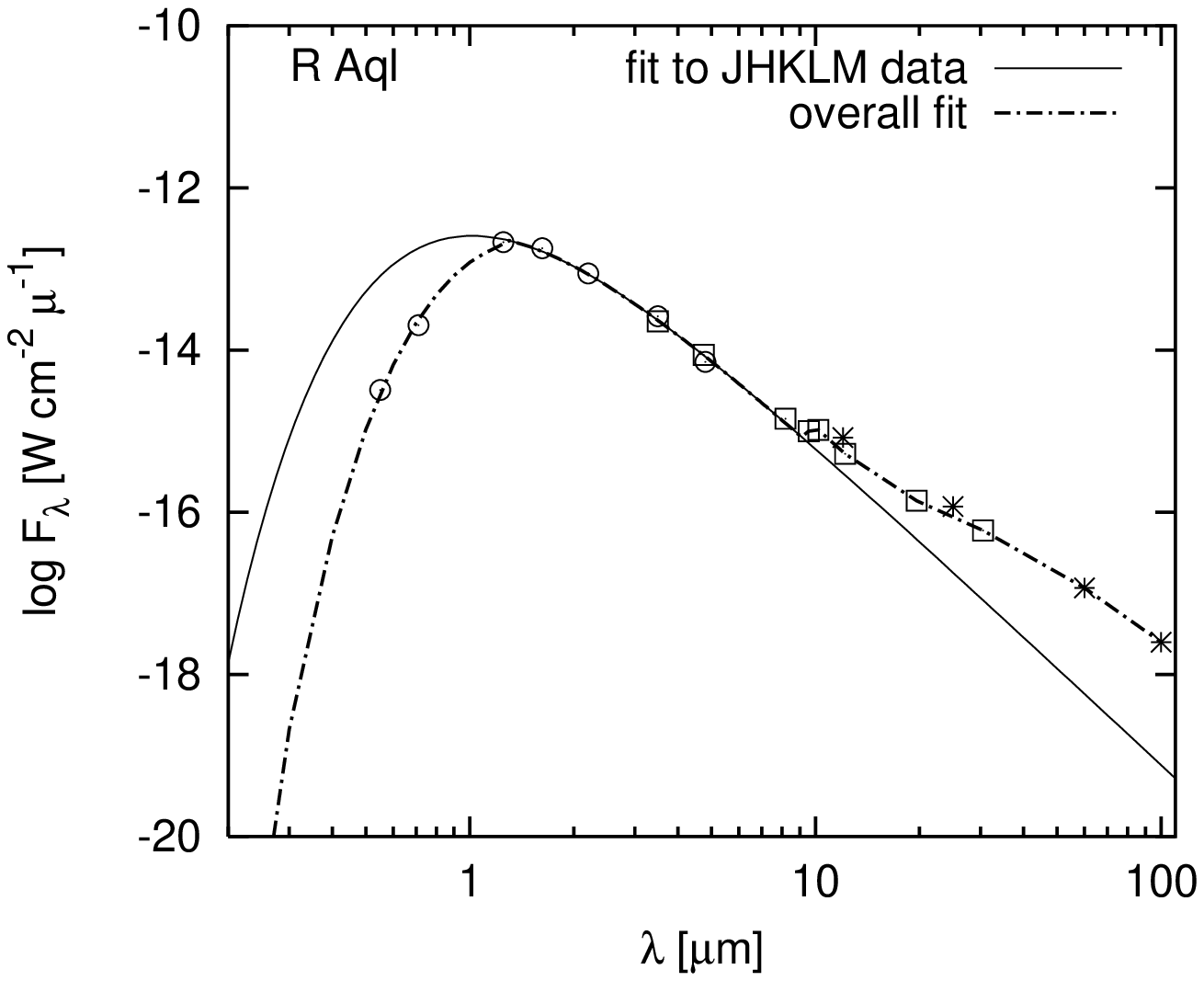}}
\epsfxsize=5.0cm
\mbox{\epsffile{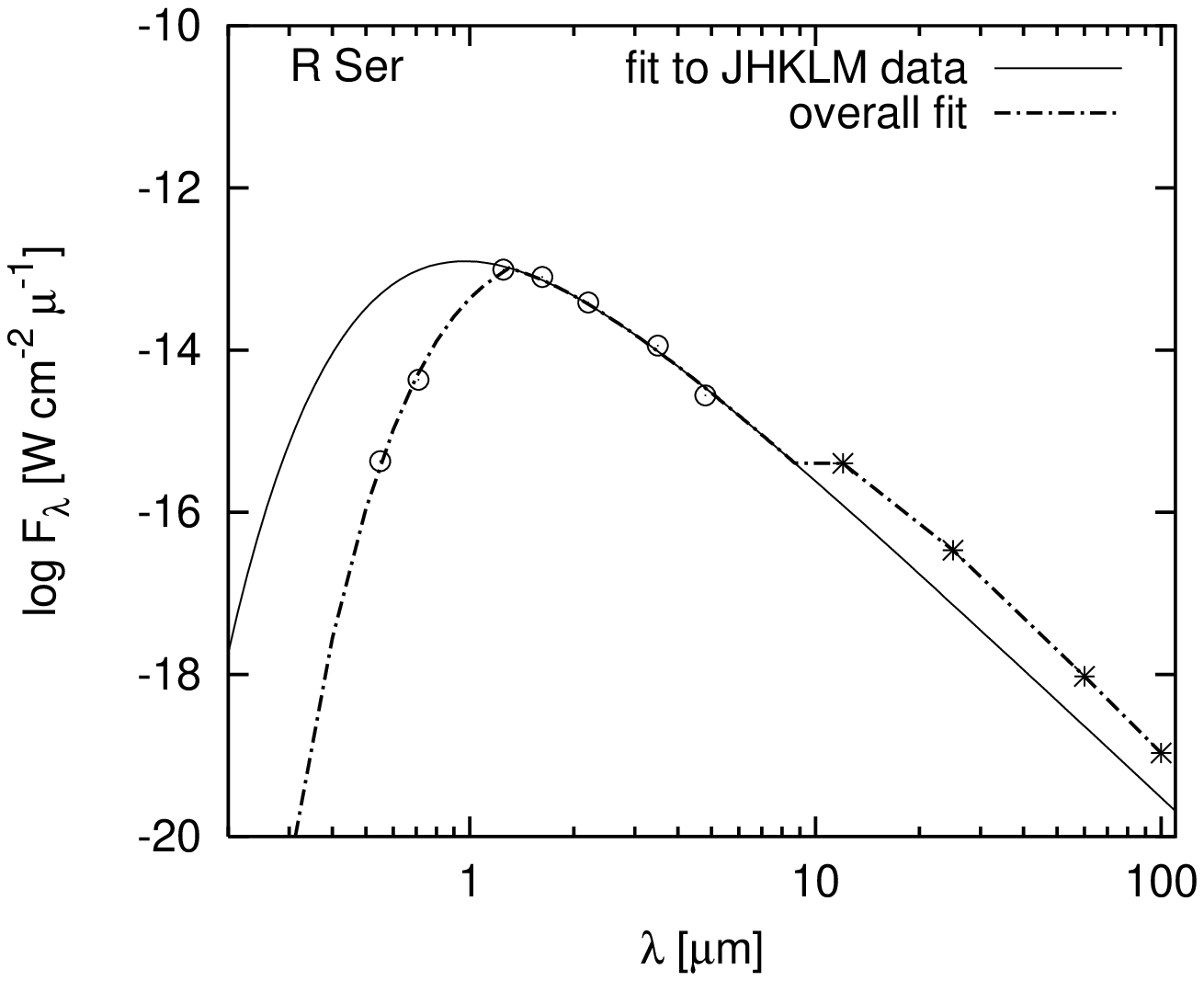}}
\caption[]{
Spectral energy distributions of \astrobj{R Aql} and \astrobj{R Ser}.
 Circles refer to AAVSO data and present photometry (see Tab.~\ref{tab:Tphoto}),
 squares to the photometry of Epchtein et al.\ (1980), and asterisks to IRAS
 (1985) data.  The solid line represents the black-body fit to the
 $JHKLM$ data, and the thick dashed line refers to the overall fit.}
   \label{fig:Fsed}
\end{figure}

The $JHKLM$ photometry can be well fitted with a black body. For the fitting
procedure the Levenberg-Marquardt method (see Press et al.\ 1992) was applied.
However, these black body fits are not well suited to represent the 
optical and mid-infrared data (see Fig.~\ref{fig:Fsed}).
In the optical regime the SED can be significantly depressed
compared to a black body due to absorption bands (e.g.\ of TiO; see
Scholz \& Takeda 1987).
For example, all oxygen-rich Mira stars of the present sample show
$V$ magnitudes well below that of a black body.
To find a fair representation of the optical flux, the $VRJ$ photometry
was fitted separately, applying the Levenberg-Marquardt method.
Fit functions resembling black bodies, but showing larger spectral indices,
proved to give good matches to the observational data
and smooth transitions to the near-infrared photometry.

On the other hand, longwards of 10\,$\mu$m dust emission may give additional,
albeit small, flux contributions. For instance, the IRAS data
of \astrobj{R Aql} does not follow the Rayleigh-Jeans curve indicating the existence
of possibly significant dust emission. However, we found 
the respective flux contributions due to dust emission to be less than 3\%
(Table~\ref{tab:Tphoto}), and therefore to be of only minor importance. 

Table~\ref{tab:Tphoto} presents the photometric data in the
$V$ ($0.55\,\mu$m, AAVSO),
$R$ ($0.71\,\mu$m, interpolated),
$J$ ($1.25\,\mu$m),
$H$ ($1.62\,\mu$m),
$K$ ($2.20\,\mu$m),
$L$ ($3.50\,\mu$m), and $M$ ($4.80\,\mu$m) band, the
pulsational phase at date of observation, the adopted extinction, and
the total bolometric flux.
The flux difference of a SED fit based only on
the $JHKLM$ photometry and the flux contribution due to dust emission are
given as well.
Fig.~\ref{fig:Fsed} illustrates photometric data and various fits for
\astrobj{R Aql} and \astrobj{R Ser}.

%

\end{document}